\documentstyle[12pt,epsf]{article}
\title{Charge on the quantum dot in the presence of tunneling
current}

\author{Mariusz Krawiec and Karol I. Wysoki\'nski\\
Institute of Physics, M. Curie-Sk\l{}odowska University,\\ 
PL-20031 Lublin, Poland} 

\begin{document} 

\maketitle
 
\begin{abstract} 
The calculation of the charge present in central region of the double
barrier structure at non-equilibrium conditions is discussed. We
propose here a simple method to calculate non equilibrium Green's
functions which allows consistent calculations of retarded and
distribution functions.  To illustrate the approach we calculate the
charge on the quantum dot coupled {\it via} tunnel barriers to two
external leads having different chemical potentials $\mu_L$ and
$\mu_R$. The obtained results have been compared  with other
approaches existing in the literature. They all agree in the
equilibrium situation and the departures grow with increasing the
difference $\mu_L-\mu_R$.

\end{abstract}

\newpage

 Recent advancements of nanotechnology allow the design and
study of devices in which quantum phenomena play primary role
\cite{meso}. Typical device consists of a small central region
coupled {\it via} tunnel barriers to two external electrodes
(leads). In such quantum-dot devices Coulomb blockade, resonant
tunneling \cite{Kouwen} and Kondo effects \cite{GGordon,SaraCr}
have been observed. Theoretical description of electron transport
phenomena in quantum dot often requires (self-consistent)
calculations of the charge accumulated in the central region
\cite{PALee,MeirWin}. If the system is in an equilibrium
state (i.e. the chemical potentials of the left and right leads are
the same $\mu_L = \mu_R = \mu$) the calculation of charge on the
dot is easy. One finds the (local) density of states for the dot
$N_d(\varepsilon)$ and integrates it with appropriate distribution
function, which in equilibrium coincides  with Fermi-Dirac one;
$f(\varepsilon) = (e^{\beta(\varepsilon-\mu)} +1)^{-1}$, with
$\beta = 1/k_BT$ and gets \begin{equation}
n = \int\limits^\infty_{-\infty} d\varepsilon
N(\varepsilon)f(\varepsilon)\,. 
\label{eq.n1}
\end{equation}
The easiest way to derive (\ref{eq.n1}) is to find  the
(equilibrium) Green's function \cite{Doniach} of the dot and apply 
spectral   theorem to get $n$. In non equilibrium situation
($\mu_L\neq \mu_R$) one can not use the equilibrium technique.
Physically the problem  is much more complicated as the
tunneling current flows across the system and there is no {\it a
priori} simple prescription as how to calculate the  charge on the
central region of the device (dot).  Some additional  complications
arise if time dependent voltage or other time dependent fields are
applied to the system.  Here, however, we shall consider 
non-equilibrium but steady state.   At nonequilibrium  one can
think about at least two ways  of generalizing Eq.
(\ref{eq.n1}). The first is to
replace the chemical potential  $\mu$ by suitably chosen average
chemical potential $\bar\mu$. The other is to replace
$f(\varepsilon)$ by the average distribution function $\bar
f(\varepsilon)$. The simplest two possibilities, which in fact have
been used in literature \cite{Haug}  to discuss various issues
connected with transport across the system at hand are the
following:  \begin{equation} \bar\mu = {\mu_L + \mu_R\over 2}
\label{eq.mubar}
\end{equation} 
or 
\begin{equation}
\bar f(\varepsilon) = {f_L(\varepsilon) +
f_R(\varepsilon)\over 2}\,,
\label{eq.fave}
\end{equation}
where $f_{L/R} = (e^{\beta(\varepsilon-\mu_{L/R})} +
1)^{-1}$, is the equilibrium distribution function for the
left/right lead. The immediate objection against the above
proposals is that both give slightly different results, except at
equilibrium and none of them is well justified. Equal weights
assigned to $\mu_{L(R)}$ or $f_{L(R)}(\varepsilon)$  may possibly be
also justified for symmetrically coupled systems only.

Below we shall present an approach which allows the correct
calculation of the charge $n$ in non equilibrium systems. To make the
presentation of it as simple as possible we shall resort
to the specific  model describing the quantum dot coupled to two
external leads. We use here the Anderson-Hubbard model
\cite{Anderson}
\begin{equation}
H = \sum_{\lambda k\sigma}(\varepsilon_{\lambda
k}-\mu_{\lambda})c^+_{\lambda k\sigma}c_{\lambda k\sigma} + E_d
\sum_\sigma d^+_\sigma d_\sigma + Un_\uparrow n_\downarrow +
\sum_{\lambda k\sigma}(V_{\lambda k} c^+_{\lambda k\sigma}d_\sigma
+ {\rm h.c.})\,. \label{eq.H} \end{equation} Here $\lambda = R, L$
denote the right ($R$) or left ($L$) lead in the system. The
parameters have the following meaning: $c^+_{\lambda
k\sigma}(c_{\lambda k\sigma})$ denote creation (annihilation)
operator for a conduction electron with wave vector $\vec k$, spin
$\sigma$ in the lead $\lambda$, $V_{\lambda k}$ is the
hybridization matrix element between conduction electron of energy
$\varepsilon_{\lambda k}$ in a lead $\lambda$ with chemical
potential $\mu_{\lambda} $ and localized electron on the dot. $E_d$
is the single particle energy of electrons in the dot.
$n_\uparrow = d^+_\uparrow d_\uparrow$ is the number operator
for electrons with spin up localized on the dot and $U$ is the
(repulsive) interaction energy between  two electrons on the dot.

The correct way to calculate the charge on the dot under  
non equilibrium conditions  is to use non equilibrium Green's
function (GF) of Keldysh \cite{Keldysh,Haug}.  
 In this technique the average charge on dot at time $t$ is given by
\begin{equation}
 \langle n(t)\rangle = \sum_\sigma\langle d^+_\sigma
(t)d_\sigma(t)\rangle=  -i\sum_\sigma G^<_\sigma(t,t)\,.
\end{equation}
where $G^<_\sigma(t,t)$ is diagonal (in time indices) element of the
Keldysh ''lesser'' GF
\begin{equation}
 G^<_\sigma(t,t') = i\langle d^+_{\sigma}(t)d_\sigma(t')\rangle\,.
\end{equation}
This function is the (1, 2) matrix element of the following contour
ordered Green's function matrix
\begin{equation}
 \widehat G(t,t') = -i\langle T_c
d(\tau)d^+(\tau')\rangle = \left(\begin{array}{ll}
 G_c, & G^< \\
 G^>, & G_{\bar c}
\end{array}\right)\,,
\end{equation}
where $T_c$ is a complex time contour ordering operator. 
The time contour starts above time axis at $t_0 = -\infty$ passes
through $t$ and $t'$ and returns back to $t_0 = -\infty$ below time
axis. $G_c = -i\langle T\,d(t)d^+(t')\rangle$ is the usual time
ordered (causal) GF for both times on the upper branch of the
contour, while $G_{\bar c}(t,t')$ is antitime ordered GF with both
time arguments on the lower branch.  Greater ($G^>$) and lesser
($G^<$) GFs are distribution functions: $G^>(t,t') = -i\langle d(t)
d^+(t')\rangle$. Not all this functions are independent. 

It turns out that the calculation of the ''lesser'' GF is easy for
noninteracting quantum dot i.e.  for $U = 0$. It can be found
directly in the following way. Because, the Keldysh contour ordered
GF, possesses the same perturbation expansion as the corresponding
equilibrium GF \cite{Keldysh} it is possible to write down the
matrix Dyson equation \begin{equation}
\widehat G = \widehat G_0 + \widehat G_0\widehat\Sigma\widehat G\,.
\end{equation}
The theorem due to Langreth \cite{Langreth} allows the transition
from contour ordered to real axis functions and one gets \cite{Haug}
\begin{equation}
 G^< = (1 + G^r\Sigma^r)\, G^<_0(1 + \Sigma^a G^a) + G^r\Sigma^< G^a
\label{eq.Keldysh}
\end{equation}
where $G^{r(a)}$ denotes retarded (advanced) GF.

For noninteracting quantum dot ($U = 0$)  exact expression for
the retarded dot GF reads
\begin{equation}
 G^r_\sigma(\omega) = G^r_{0\sigma}(\omega) + G^r_{0\sigma}(\omega) \,
 \Sigma^r(\omega)\, G^r_\sigma(\omega)
\label{eq.Dysonr}
\end{equation}
with
\begin{equation}
\Sigma^r(\omega) = \sum_{\lambda k} |V_{\lambda k}|^2 G^r_{0\lambda
k}(\omega) \,,
\label{eq.selfr}
\end{equation}
where $G^r_{0\lambda k}(\omega) = (\omega - \varepsilon_{\lambda k} +
i0)^{-1}$ is the Green's function of the lead $\lambda$. From this
one immediately finds  
\begin{eqnarray}
\Sigma^<(\omega) &=& \sum_{\lambda k} {V_{\lambda k}}^2 G^<_{0\lambda
k} (\omega) =  i\sum_{\lambda k} |V_{\lambda k}|^2 \delta(\omega - 
\varepsilon_{\lambda k}) f_\lambda(\omega) = \nonumber \\
&=& 2\pi i \sum_\lambda\Gamma_\lambda(\omega) f_\lambda (\omega)\,,
\label{eq.Sigl}
\end{eqnarray}
where we have used symbol $\Gamma_\lambda(\omega)$ to denote the
average coupling of the dot to lead $\lambda$; 
$\Gamma_\lambda(\omega) = 2\pi \sum_k|V_{\lambda
k}|^2\delta(\omega-\varepsilon_{\lambda k})$. Noting that $G^<(t,t)
= \int{d\omega\over 2\pi}G^<(\omega)$ and using formula
(\ref{eq.Keldysh}) we get  \begin{equation}
\langle n\rangle = \sum_\sigma\int d\omega\,G^r(\omega)
\Sigma^<(\omega)  G^a(\omega)\, ,
\end{equation}
which with help of Eqs. (\ref{eq.Dysonr}) and (\ref{eq.Sigl}) we
further rewrite  as 
\begin{equation}
n = \sum_\sigma\int d\omega {\sum\limits_\lambda
\Gamma_\lambda(\omega)f_\lambda(\omega) \over
\sum\limits_\lambda\Gamma_\lambda(\omega)} \left(-{1\over\pi}\right) {\rm Im}
G^r_\sigma(\omega)\,.
\end{equation}
Note that for symmetric coupling $\Gamma_L(\omega) =
\Gamma_R(\omega)$ this formula reduces to the form 
(\ref{eq.n1}) 
with average
distribution function $\bar f(\varepsilon)$ as given by 
(\ref{eq.fave}). 
Such an expression
for dot occupation has been used in \cite{SunLin} for an interacting
dot. We stress, however,  that it is valid for noninteracting and
symmetrically coupled dot only.  

As one can see from the above the main difficulty in calculation of
the charge on the dot is connected with calculation of the
''lesser'' self energy $\Sigma^<$. In the noninteracting case the
knowledge of the exact expression for self energy (\ref{eq.selfr})
allowed easy calculation of exact $\Sigma^<$ and exact expression
for the charge. For interacting dot there is no way to get
self energy exactly and one has to approximate it appropriately. One
such approximation has been introduced by Ng \cite{Ng} and will be
discussed latter on. Here we  propose to use recently derived
equation of motion method \cite{Niu} and calculate $G^<(\omega)$
with the desired accuracy and with approximations consistent with
those made in calculation of the current or other response
functions of the system.

To make the algebra easy we consider $U =
\infty$ limit in the Hamiltonian (\ref{eq.H}) and use slave boson
technique to handle it. We thus rewrite (\ref{eq.H}) with help of
slave boson operators $b, b^+$ and use the commutation rules of
LeGuillou and Ragoucy \cite{LeGRag} to evaluate the necessary quantum
brackets. These rules treat exactly the local constraint
\cite{PRB99} which for $U=\infty$ prevents double
occupancy of the dot. The procedure is simple. One rewrites the
Hamiltonian in the form  
\begin{equation}
H^{SB} = \sum_{\lambda k\sigma}(\varepsilon_{\lambda
k}-\mu_{\lambda}) c^+_{\lambda k\sigma} c_{\lambda k\sigma} +
\varepsilon_d \sum_\sigma f^+_\sigma f_\sigma + \sum_{\lambda
k\sigma} V_{\lambda k} (c^+_{\lambda k\sigma} b^+f_{\sigma} +
f^+_\sigma b c_{\lambda k\sigma}) 
\end{equation}
and calculates the on-dot Green's function $D^<_\sigma(\omega) =
\langle\langle b^+f_\sigma |f^+_\sigma b\rangle\rangle^<_\omega$ 
using equation ({\it c.f.} equation (28b) of [8])
\begin{equation}
\langle\langle A|B\rangle\rangle =
g^<(\omega)\langle[A,B]_{\pm}\rangle+g^r(\omega)
\langle\langle[A,H_I]|B\rangle\rangle^<_{\omega}+
g^<(\omega)\langle\langle[A,H_I]|B\rangle\rangle^a_{\omega},
\end{equation}
with the third term of $H^{SB}$ taken as a
interaction part $H_I$ and lower case GFs being the corresponding
GFs of the free Hamiltonian $H_0$ (consisting of first and second
terms of $H^{SB}$). The higher order GFs appearing at this stage
have been calculated in similar way. In the process we have used
the same kind of factorization which one uses calculating the
on-dot retarded GF necessary to get Kondo effect.  Explicitly we
neglected the GFs like $<<c_{\lambda k-\sigma}c_{\lambda'k'\sigma}
f^+_{-\sigma}b|f^+_\sigma b>>^a $ as they describe higher order
spin correlations in the leads and performed the decoupling 
\begin{equation}
<<c^+_{\lambda'k'-\sigma} c_{\lambda k-\sigma} b^+f_{\sigma}
|  f^+_\sigma b>>^a \approx f(\varepsilon_{\lambda
k}) << b^+f_\sigma|f^+_\sigma b>>^a \delta_{kk'}
\delta_{\lambda\lambda'}\,. \end{equation}
 The resulting ''lesser'' self-energy takes on simple form 
\begin{equation}
\Sigma^<(\omega) = -2\pi i \sum_\lambda\Gamma_\lambda(\omega) (1 +
f_\lambda(\omega))f_\lambda(\omega)  \,.
\label{eq.ourSig}
\end{equation}
This is our final expression for $\Sigma^<(\omega)$ to be used
in calculations of the charge $n$.

Now let us rederive the expression for
$\Sigma^<(\omega)$ using the earlier mentioned approximation
introduced by Ng \cite{Ng} and often used in the literature
\cite{Raimondi}.  This author proposes to assume that 
$\Sigma^<(\omega) = A\Sigma^<_0(\omega)$ and $\Sigma^>(\omega) =
A\Sigma^>_0$ where $A$ is an unknown function and 
$\Sigma^<_0(\omega)$ is noninteracting (and thus  known) self-energy.
For the present example it is given explicitly by (\ref{eq.Sigl}).
This together with the exact relations: $\Sigma^> - \Sigma^< =
\Sigma^r - \Sigma^a$ and $\Sigma^>_0 - \Sigma^<_0 = \Sigma^r_0 -
\Sigma^a_0$ leads to  \begin{equation} \Sigma^<_{\rm
Ng}(\omega)={\Sigma^r - \Sigma^a\over\Sigma^r_0 - \Sigma^a_0}
\Sigma^<_0(\omega) \end{equation}
  or in explicit form 
\begin{equation}
\Sigma_{\rm Ng}^<(\omega) = -2\pi i
{\sum\limits_\lambda\Gamma_\lambda(\omega) f_\lambda(\omega)\over
\sum\limits_\lambda \Gamma_\lambda(\omega)} \, \sum_\lambda
\Gamma_\lambda (1 + f_\lambda(\omega))\,,
\label{eq.Ng}
\end{equation}
which again contains the characteristic average distribution
function, which for the symmetric coupling reduces to $\bar
f(\varepsilon)$.  Note that this formula agrees with
(\ref{eq.ourSig}) only in equilibrium situation $\mu_L = \mu_R =
\mu$.

It turns out that our formula, Eq. (\ref{eq.ourSig}) can be rederived
by suitably generalizing the Ng's {\it ansatz}. To see this note
that noninteracting self energy $\Sigma^<_0(\omega)$ above can be
written as a sum of pieces, each of which is connected with
one of the leads i.e. $\Sigma^<_0(\omega) = \Sigma_\lambda
\Sigma^<_{0\lambda}(\omega)$. One then expects that due to locality
of $U$ term in Hamiltonian the interacting self-energy will also be
a sum of contributions from different leads  $\Sigma^<(\omega) =
\sum_\lambda \Sigma^<_\lambda(\omega)$. Generalising the {\it
ansatz} and writing $\Sigma^<_\lambda(\omega) =
A_\lambda\Sigma^<_{0\lambda}$ independently for each
$\lambda$ one immediately reproduces result
(\ref{eq.ourSig}). This shows that the application of equation of
motion technique \cite{SunLin} is a correct way to to derive the
"lesser" self energy. This technique, contrary to  approximate
schemes, allows mutually consistent calculations of both retarded and
distribution GFs in Keldysh technique. 

\begin{figure}[ht] 
\epsfysize=5cm  
\centerline{\epsfbox{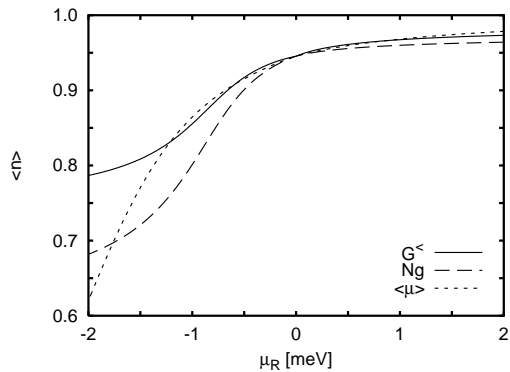}} 
\caption{The charge on the  quantum dot connected to two external 
electrodes in dependence of the chemical potential $\mu_R$.  The 
parameters are the following: couplings $\Gamma_L=\Gamma_R= 17
meV$, chemical potential $\mu_L = 0$   and temperature $T =
10^{-4}meV/k_B $.}     
\label{fig.nd1} 
\end{figure} 

 We have numerically calculated the charge $n$ on the interacting 
quantum dot (with $U=\infty$) coupled to leads with different
chemical potentials using four formulae discussed in this
paper. The results are presented  in figures (1) and (2). Figure (1) shows
the charge on the dot  with $E_d=-1 meV$ as a function of the
position of right lead chemical potential $\mu_R$ with $\mu_L=0$ and
for symmetric couplings $\Gamma_L(\omega)=\Gamma_R(\omega)=\Gamma =
17 meV$. Note tha for symmetric coupling the Ng's approximation give
the same results as average distribution function $\bar
f(\varepsilon)$ . In more general case of asymmetric coupling these
two approaches lead to different results, but  we abandoned this
complication here.  Solid line show results obtained with help of our
formula (\ref{eq.ourSig}), dashed one is calculated with help of
(\ref{eq.Ng}),  while dotted from (\ref{eq.n1}) with $\bar \mu$
given by (\ref{eq.mubar}). The crossing of the lines for
$\mu_L=\mu_R=0$ corresponds to equilibrium situation. The
differences between the curves get larger for increasing voltage
$V=(\mu_L-\mu_R)/e$ applied to the system. The large
(small) differences between various curves for negative (positive)
values of $\mu_R$ is due to large (small) values of
the on-dot density of states in the energy window between $\mu_L$
and $\mu_R$.
\begin{figure}[ht] 
\epsfysize=5cm   
\centerline{\epsfbox{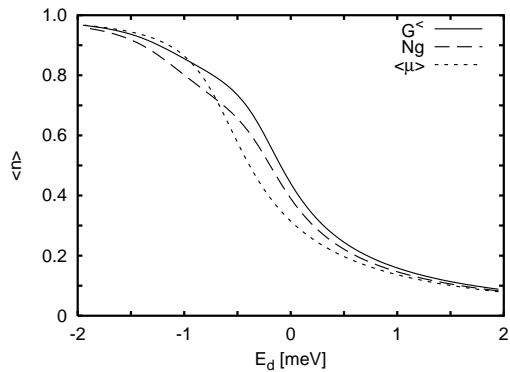}} 
\caption{The charge  $n$ {\it vs} $E_d$  for constant difference 
$\mu_R -\mu_L= 1 meV$. Other  parameters are the same as in
Fig.(\ref{fig.nd1}).}   
\label{fig.nd2} 
\end{figure} 
Figure (2) shows the charge on the dot calculated for constant
difference $\mu_L-\mu_R=1 meV$ as a function of the parameter $E_d$.
For $E_d$ much higher or much lower than both chemical potentials the
differences are small as expected because the density of states
around $\mu_L$ and $\mu_R$ is small. The largest differences of the
calculated charge are found for such values of $E_d$ for which the
density of states of electrons on the dot possesses the Kondo
resonance. 

We conclude by stressing that the calculation of charge on the dot
is a necessary step towards self consistent calculations of the
current flowing across quantum dot devices. In some parameter ranges
all formulae give only slightly differing results. Generally,
however, the differences between   values of $n$ calculated by means
of different formulae may be as large as 20-30\%. The equation of
motion method allows calculation of the lesser self energy
consistent with retarded one. This is important as the current
flowing in the system is, for proportionate couplings
$\Gamma_L(\omega)=const \Gamma_R(\omega)$, expressed by the retarded 
Green's function which in turn depends on the charge $n$ on the dot.
For general couplings the consistency of approximations is even more
important as the current depends on both retarded and lesser GFs and
to be consistent one has to treat both self energies on equal
footing.

{\it This work has been partially supported by the Committee for
Scientific  Research under grant 2P03B 106 17. }

\end{document}